\RequirePackage{lineno}
\documentclass[aps,prl,twocolumn,showpacs,superscriptaddress,floatfix]{revtex4}
\usepackage{epsfig}
\setpagewiselinenumbers

\newcommand {\snn}	{\sqrt{s_{_{\rm NN}}}}
\newcommand {\gev}	{{GeV/$c$}}

\newcommand {\pp}	{{$p$+$p$}}
\newcommand {\ppb}	{{$p$+Pb}}
\newcommand {\dAu}	{{$d$+Au}}

\newcommand {\zyam}	{ZYAM}
\newcommand {\hijing}	{HIJING}
\newcommand {\pt}	{p_{T}}
\newcommand {\ptt}	{p_{T}^{(t)}}
\newcommand {\pta}	{p_{T}^{(a)}}

\newcommand {\dphi}	{\Delta\phi}
\newcommand {\deta}	{\Delta\eta}
\newcommand {\Ntrig}	{N_{\rm trig}}
\newcommand {\Yjet}	{Y_{\rm jetlike}}

\newcommand {\note}[1]	{}

\begin{document}
\title{Effect of event selection on jetlike correlation measurement in \dAu\ collisions at $\snn=200$~GeV }
\affiliation{AGH University of Science and Technology, Cracow 30-059, Poland}
\affiliation{Argonne National Laboratory, Argonne, Illinois 60439, USA}
\affiliation{Brookhaven National Laboratory, Upton, New York 11973, USA}
\affiliation{University of California, Berkeley, California 94720, USA}
\affiliation{University of California, Davis, California 95616, USA}
\affiliation{University of California, Los Angeles, California 90095, USA}
\affiliation{Universidade Estadual de Campinas, Sao Paulo 13131, Brazil}
\affiliation{Central China Normal University (HZNU), Wuhan 430079, China}
\affiliation{University of Illinois at Chicago, Chicago, Illinois 60607, USA}
\affiliation{Creighton University, Omaha, Nebraska 68178, USA}
\affiliation{Czech Technical University in Prague, FNSPE, Prague, 115 19, Czech Republic}
\affiliation{Nuclear Physics Institute AS CR, 250 68 \v{R}e\v{z}/Prague, Czech Republic}
\affiliation{Frankfurt Institute for Advanced Studies FIAS, Frankfurt 60438, Germany}
\affiliation{Institute of Physics, Bhubaneswar 751005, India}
\affiliation{Indian Institute of Technology, Mumbai 400076, India}
\affiliation{Indiana University, Bloomington, Indiana 47408, USA}
\affiliation{Alikhanov Institute for Theoretical and Experimental Physics, Moscow 117218, Russia}
\affiliation{University of Jammu, Jammu 180001, India}
\affiliation{Joint Institute for Nuclear Research, Dubna, 141 980, Russia}
\affiliation{Kent State University, Kent, Ohio 44242, USA}
\affiliation{University of Kentucky, Lexington, Kentucky, 40506-0055, USA}
\affiliation{Korea Institute of Science and Technology Information, Daejeon 305-701, Korea}
\affiliation{Institute of Modern Physics, Lanzhou 730000, China}
\affiliation{Lawrence Berkeley National Laboratory, Berkeley, California 94720, USA}
\affiliation{Massachusetts Institute of Technology, Cambridge, Massachusetts 02139-4307, USA}
\affiliation{Max-Planck-Institut fur Physik, Munich 80805, Germany}
\affiliation{Michigan State University, East Lansing, Michigan 48824, USA}
\affiliation{Moscow Engineering Physics Institute, Moscow 115409, Russia}
\affiliation{National Institute of Science Education and Research, Bhubaneswar 751005, India}
\affiliation{Ohio State University, Columbus, Ohio 43210, USA}
\affiliation{Institute of Nuclear Physics PAN, Cracow 31-342, Poland}
\affiliation{Panjab University, Chandigarh 160014, India}
\affiliation{Pennsylvania State University, University Park, Pennsylvania 16802, USA}
\affiliation{Institute of High Energy Physics, Protvino 142281, Russia}
\affiliation{Purdue University, West Lafayette, Indiana 47907, USA}
\affiliation{Pusan National University, Pusan 609735, Republic of Korea}
\affiliation{University of Rajasthan, Jaipur 302004, India}
\affiliation{Rice University, Houston, Texas 77251, USA}
\affiliation{University of Science and Technology of China, Hefei 230026, China}
\affiliation{Shandong University, Jinan, Shandong 250100, China}
\affiliation{Shanghai Institute of Applied Physics, Shanghai 201800, China}
\affiliation{SUBATECH, Nantes 44307, France}
\affiliation{Temple University, Philadelphia, Pennsylvania 19122, USA}
\affiliation{Texas A\&M University, College Station, Texas 77843, USA}
\affiliation{University of Texas, Austin, Texas 78712, USA}
\affiliation{University of Houston, Houston, Texas 77204, USA}
\affiliation{Tsinghua University, Beijing 100084, China}
\affiliation{United States Naval Academy, Annapolis, Maryland, 21402, USA}
\affiliation{Valparaiso University, Valparaiso, Indiana 46383, USA}
\affiliation{Variable Energy Cyclotron Centre, Kolkata 700064, India}
\affiliation{Warsaw University of Technology, Warsaw 00-661, Poland}
\affiliation{Wayne State University, Detroit, Michigan 48201, USA}
\affiliation{World Laboratory for Cosmology and Particle Physics (WLCAPP), Cairo 11571, Egypt}
\affiliation{Yale University, New Haven, Connecticut 06520, USA}
\affiliation{University of Zagreb, Zagreb, HR-10002, Croatia}

\author{L.~Adamczyk}\affiliation{AGH University of Science and Technology, Cracow 30-059, Poland}
\author{J.~K.~Adkins}\affiliation{University of Kentucky, Lexington, Kentucky, 40506-0055, USA}
\author{G.~Agakishiev}\affiliation{Joint Institute for Nuclear Research, Dubna, 141 980, Russia}
\author{M.~M.~Aggarwal}\affiliation{Panjab University, Chandigarh 160014, India}
\author{Z.~Ahammed}\affiliation{Variable Energy Cyclotron Centre, Kolkata 700064, India}
\author{I.~Alekseev}\affiliation{Alikhanov Institute for Theoretical and Experimental Physics, Moscow 117218, Russia}
\author{J.~Alford}\affiliation{Kent State University, Kent, Ohio 44242, USA}
\author{A.~Aparin}\affiliation{Joint Institute for Nuclear Research, Dubna, 141 980, Russia}
\author{D.~Arkhipkin}\affiliation{Brookhaven National Laboratory, Upton, New York 11973, USA}
\author{E.~C.~Aschenauer}\affiliation{Brookhaven National Laboratory, Upton, New York 11973, USA}
% ...

\collaboration{STAR Collaboration}\noaffiliation

\begin{abstract}
     Dihadron correlations are analyzed in $\sqrt{s_{_{\rm NN}}} = 200$ GeV \dAu\ collisions classified by forward charged particle multiplicity and zero-degree neutral energy in the Au-beam direction. It is found that the jetlike correlated yield increases with the event multiplicity. After taking into account this dependence, the non-jet contribution on the away side is minimal, leaving little room for a back-to-back ridge in these collisions.
\end{abstract}
\maketitle


High transverse momentum ($\pt$) particle yield measured at the Relativistic Heavy Ion Collider (RHIC) was found to be strongly suppressed in relativistic heavy-ion collisions compared to elementary proton-proton collisions~\cite{Arsene:2004fa,Back:2004je,Adams:2005dq,Adcox:2004mh}. 
It was concluded that the strong high-$\pt$ suppression was due to final-state effects in the hot and dense quark-gluon plasma created in those collisions~\cite{Arsene:2004fa,Back:2004je,Adams:2005dq,Adcox:2004mh}.
Instrumental to this conclusion was the control experiment of proton-nucleus, or deuteron-gold (\dAu) collisions as realized at RHIC, that excluded cold nuclear effects as the possible primary cause for the suppression~\cite{Arsene:2004fa,Back:2004je,Adams:2005dq,Adcox:2004mh}. 
The observations of the long-range pseudorapidity separation ($\deta$) dihadron correlations at small relative azimuth ($\dphi$) in control experiments 
 \pp\ and \ppb~\cite{CMS:2012qk,Abelev:2012ola,Aad:2012gla} collisions at the Large Hadron Collider (LHC) were therefore surprising, because the observed long-range correlations were similar to the novel long-range correlation first discovered in heavy-ion collisions at RHIC~\cite{Adams:2005ph,Abelev:2009af,Alver:2009id,Abelev:2009jv}, called the ``ridge.'' The heavy-ion ridge was primarily attributed to collective anisotropic flow~\cite{Alver:2010gr}. Collective flow is not normally expected for small collision systems where the dihadron correlations are dominated by jet correlations. 
To reduce or remove jet contributions, 
dihadron correlation in low-multiplicity collisions was subtracted from that in 
high-multiplicity collisions in previous experiments~\cite{Abelev:2012ola,Aad:2012gla,Adare:2013piz}. 
Applying such a subtraction procedure 
revealed 
a back-to-back ridge at $\dphi\sim\pi$, along with the ridge at $\dphi\sim0$ in \ppb\ at $\snn=5.02$~TeV~\cite{Abelev:2012ola,Aad:2012gla}. 
Using the same subtraction technique, PHENIX also observed a (near- and away-side) double ridge in \dAu\ collisions at $\snn=200$~GeV within 
$|\deta|<0.7$~\cite{Adare:2013piz}. 
As observed in larger systems, the double ridge is reminiscent of a non-jet elliptic flow contribution~\cite{Bozek:2010pb,Bozek:2012gr}. 
Other physics mechanisms 
have however also been proposed, 
such as the color glass condensate where two-gluon densities are enhanced at small $\dphi$ over a wide range of $\deta$~\cite{Dumitru:2010iy,Dusling:2012wy,Dusling:2013oia}, or quantum initial anisotropy from the space momentum uncertainty principle~\cite{Molnar:2014mwa}. 

The difference in dihadron correlations between high- and low-multiplicity events would be attributable to non-jet physics if jetlike correlations are identical in these two event classes.
However, since jet particle production contributes to the overall multiplicity, 
the selection of high-multiplicity events may demand a relatively large number of jet-correlated particles. 
In fact, such differences have been observed previously by the STAR experiment in two-particle correlations in \pp\ and various multiplicity \dAu\ collisions~\cite{Adams:2004bi,Adams:2003im}.
Most studies to date have attempted to remove/reduce the simple auto-correlations between jet production and enhanced multiplicity by selecting events via multiplicity measurements at large $\deta$ from the jet.
STAR, with its pseudorapidity and azimuthal coverage larger than typical jet sizes, is well suited to investigate the details of dihadron jetlike correlations and possible effects from event selection.


The data reported here were taken during the \dAu\ run in 2003 by the STAR experiment~\cite{Adams:2003im,Abelev:2008ab}. 
The details of the STAR experiment can be found in Ref.~\cite{Ackermann:2002ad}. 
Minimum-bias (MB) \dAu\ events were triggered by coincidence of signals from the Zero Degree Calorimeters (ZDC) $|\eta|>6.5$~\cite{Adler:2003sp} and the Beam-Beam Counters (BBC)~\cite{Ackermann:2002ad}. 
Charged particle tracks were reconstructed in the Time Projection Chamber (TPC)~\cite{Anderson:2003ur} and the forward TPC (FTPC)~\cite{Ackermann:2002yx}.
The primary vertex was determined from reconstructed tracks in the TPC.
In this analysis events were required to have a primary vertex position $|z_{\rm vtx}|<50$~cm from the  center of TPC. 
Particle tracks used in the correlation analysis were from the TPC ($|\eta|<1$), and 
required to have at least 25 out of the maximum possible of 45 hits and a distance of closest approach to the primary vertex within 3~cm.

Two quantities were used to select \dAu\ events: the charged particle multiplicity within $-3.8<\eta<-2.8$ measured by the FTPC in the Au-beam direction (FTPC-Au)~\cite{Adams:2003im,Abelev:2008ab} 
and the neutral energy (attenuated ADC signal) measured by the ZDC in the Au-beam direction (ZDC-Au). 
These measures are referred to, in this article, generally as ``event activity.''
While positive but weak correlations were observed between these measures,
the same event fraction percentage defined by these measures, e.g.~events with the 0-20\% highest FTPC-Au multiplicities or ZDC-Au energies, correspond to significantly different \dAu\ event samples.

The two particles in pairs used in dihadron correlations are customarily called trigger and associated particle~\cite{Adams:2005dq}. 
The trigger particle is typically chosen at high $\pt$ and all other particles are used as associated particles.
In this analysis pair density distributions $\frac{1}{\Ntrig}\frac{d^2N}{d\deta d\dphi}$ are measured in relative azimuthal angle $\dphi$ and pseudorapidity distance $\deta$ 
and are normalized by the number of trigger particles.
The correlation data are corrected for the associated particle tracking efficiency of $85\%\pm5\%$~(syst.)~\cite{Abelev:2008ab,Adams:2003im}, which does not vary from low to high event activity in \dAu\ collisions. 
Here, high(low) event activity 
refers to event classes selected by high(low) FTPC-Au multiplicities or ZDC-Au neutral energies. 
%
The detector non-uniformity in $\dphi$ and acceptance in $\deta$ is corrected by the event-mixing technique, where the trigger particle from one event is paired with associated particles from another event. To reduce statistical fluctuations, each trigger particle is mixed with associated particles from ten other events. 
The mixed events are required to be within 1~cm in $z_{\rm vtx}$, with the same multiplicity (measured by FTPC-Au) or within similar zero-degree neutral energy (measured by ZDC-Au). The mixed-event correlations are normalized to 100\% at $\deta=0$.

\begin{figure}
\begin{center}
\includegraphics[width=0.555\textwidth]{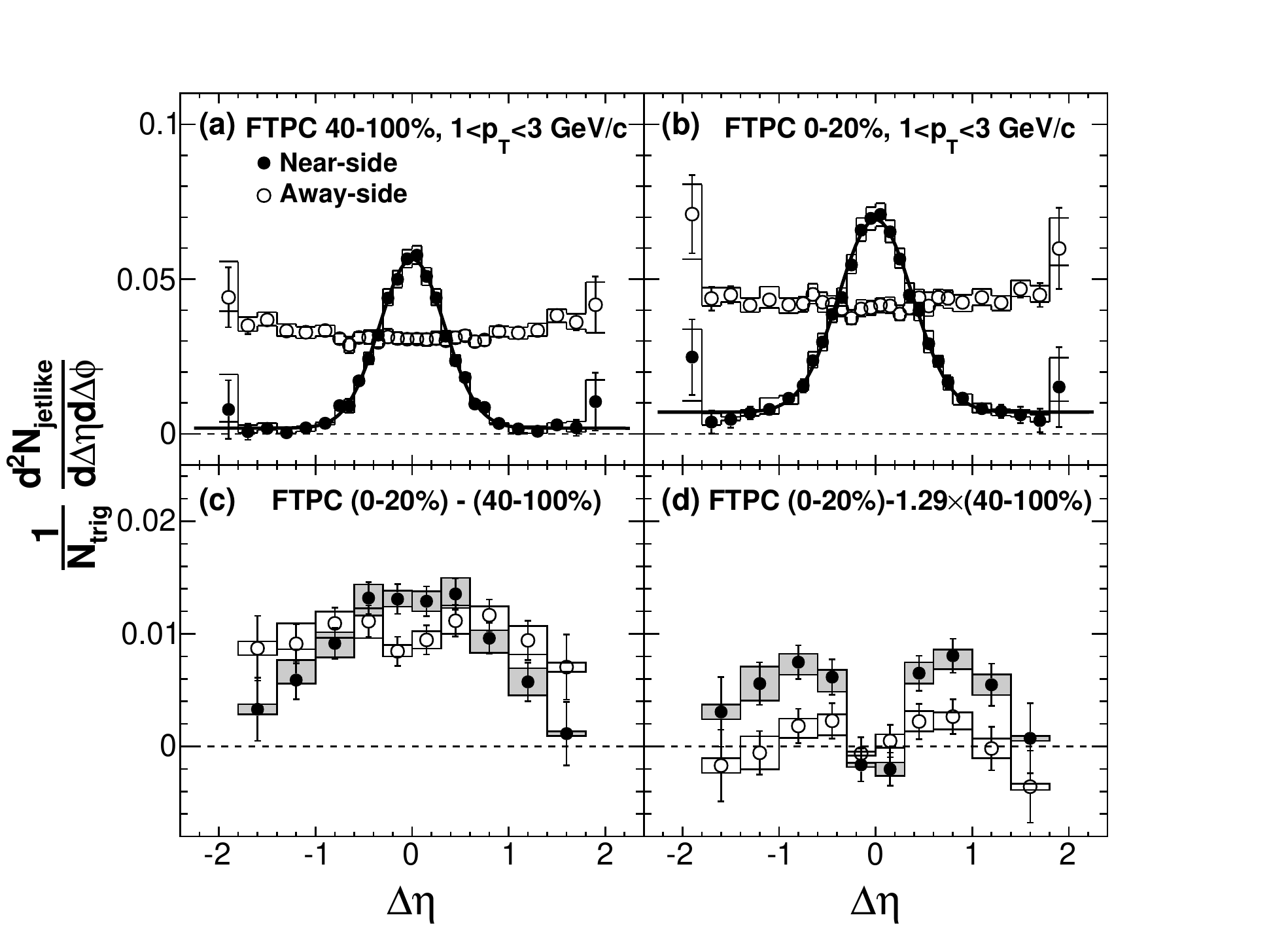}
\end{center}
\caption{The dihadron correlated yield normalized per radian per unit of pseudorapidity as function of $\deta$ in \dAu\ collisions on the near ($|\dphi|<\pi/3$, solid circles) and away side ($|\dphi-\pi|<\pi/3$, open circles). Shown are the (a) low and (b) high FTPC-Au activity data, and the high-activity data after subtracting the (c) unscaled and (d) scaled low-activity data. Trigger and associated particles have $1<\pt<3$~\gev\ and $|\eta|<1$. The Gaussian$+$pedestal fit to the near side is
superimposed as the solid curves. Error bars are statistical and boxes indicate the systematic uncertainties.}
\label{fig:deta}
\end{figure}

Dihadron correlations, after combinatorial background subtraction, are often used to study correlations originating from jets~\cite{Adams:2005dq}. However, other correlations than jets are also present, such as resonance decays. The parts of the dihadron correlations used for the jet study are therefore referred to as ``jetlike'' correlations in this Letter. 
In order to obtain jetlike correlations in \dAu\ collisions, a uniform combinatorial background is subtracted. The background normalization is estimated by the Zero-Yield-At-Minimum (\zyam) assumption~\cite{Adams:2005ph,Ajitanand:2005jj}.   
After the correlated yield distribution is folded into the range of $0<\dphi<\pi$, \zyam\ is taken as the lowest yield average over a $\dphi$ window of $\pi/8$ radian width. The \zyam\ systematic uncertainty is estimated by the yields at the \zyam\ $\dphi$ location averaged over ranges of width of $\pi/16$ and $3\pi/16$ radians. We also fit the $\Delta\phi$ correlations by two Gaussians (with centroids fixed at 0 and $\pi$) plus a pedestal. The fitted pedestal is consistent with ZYAM within the statistical and systematic errors because the near- and away-side peaks are well separated in $d$+Au collisions. 
Figure~\ref{fig:deta}(a) and \ref{fig:deta}(b) show the correlated yield densities per radian per unit of pseudorapidity as a function of $\deta$ for both the near-side ($|\dphi|<\pi/3$) and away-side ($|\dphi-\pi|<\pi/3$) ranges in (a) low and (b) high FTPC-Au multiplicity collisions. 
Both the trigger and associated particle $\pt$ ranges are $1<\pt<3$~\gev.
The \zyam\ background estimate is done for individual $\deta$ bins separately. The statistical errors of the data points include point-to-point statistical errors from the \zyam\ values, since each $\deta$ bin has its own ZYAM value. 
The near-side yields exhibit Gaussian peaks and the away-side yields are approximately uniform in $\deta$. A Gaussian+pedestal function  $\frac{\Yjet}{\sqrt{2\pi}\sigma}\exp\left(-\frac{(\deta)^2}{2\sigma^2}\right)+C$ fits to the near-side data are superimposed in Fig.~\ref{fig:deta}(a,b) as solid curves, and the fit parameters are listed in Table~\ref{tab:yields}. The Gaussian area $\Yjet$ measures the near-side jetlike correlated yield per radian. The fits
indicate a ratio   
$\alpha=\Yjet^{\rm high}/\Yjet^{\rm low}=1.29 \pm 0.05({\rm stat.}) \pm 0.02({\rm syst.})$  
of jetlike yields in high to low FTPC-Au multiplicity collisions. For ZDC-Au event selection, the jetlike ratio parameter is 
$\alpha=1.13 \pm 0.05({\rm stat.}) \pm 0.03({\rm syst.})$. 
The $\alpha$ parameter for events selected by FTPC-Au multiplicity is further from unity compared to $\alpha$ for events selected by ZDC-Au energy.
The ratios of the away-side correlated yields are 
$1.32 \pm 0.02({\rm stat.}) \pm 0.01({\rm syst.})$ for FTPC-Au multiplicity 
and 
$1.22 \pm 0.02({\rm stat.}) \pm 0.01({\rm syst.})$ for ZDC-Au energy selected events 
respectively. 
The correlated yield ratios are similar (within 2 standard deviations) between the near and away side, consistent with back-to-back jet correlations. 
In addition, the near-side Gaussian peak is wider in high- than in low-activity collisions.
A similar broadening of jetlike peak was previously observed in \dAu\ collisions compared with that in \pp\ collisions~\cite{Adams:2003im}.
\begin{table}
\caption{Gaussian+pedestal $\frac{\Yjet}{\sqrt{2\pi}\sigma}\exp\left(-\frac{(\deta)^2}{2\sigma^2}\right)+C$ fit results to near-side correlated yield densities in \dAu\ collisions. The percentiles indicate fractions of selected events, 40-100\% being low-activity and 0-20\% high-activity. First errors are statistical, and second systematic (due to \zyam). An additional 5\% efficiency uncertainty applies to $\Yjet$ and $C$.}
\label{tab:yields}
\begin{tabular}{lr|cccc}\hline
\multicolumn{2}{c|}{Event selection} & $\chi^2/{\rm ndf}$ & $\sigma(\times10^{-3})$ & $\Yjet(\times10^{-4})$ & $C(\times10^{-4})$ \\\hline
%
FTPC & 40-100\% & 19/25 & 336$\pm$7$\pm$1 & 461$\pm$11$\pm$5 & 19$\pm$5$\pm$9 \\
     & 20-40\%  & 18/25 & 362$\pm$8$\pm$3 & 546$\pm$$15^{+7}_{-14}$ & 24$\pm$$7^{+20}_{-11}$ \\
     & 0-20\%   & 19/25 & 382$\pm$10$\pm$9 & 596$\pm$$19^{+15}_{-11}$ & 70$\pm$8$\pm$12 \\\hline
ZDC  & 40-100\% & 19/25 & 352$\pm$$7^{+2}_{-6}$ & 501$\pm$11$\pm$1 & 22$\pm$$5^{+14}_{-8}$ \\
     & 20-40\%  & 26/25 & 372$\pm$9$\pm$7 & 580$\pm$18$\pm$17 & 43$\pm$8$\pm$12 \\
     & 0-20\%   & 17/25 & 376$\pm$10$\pm$3 & 568$\pm$20$\pm$17 & 59$\pm$$9^{+27}_{-14}$ \\\hline
%
\end{tabular}
\end{table}

In previous studies, dihadron correlations in low-multiplicity events are subtracted from high-multiplicity events. The residual correlation is often attributed to non-jet origins assuming jetlike correlations are equal in high- and low-multiplicity collisions~\cite{Adare:2013piz}.
The differences between 
high and low FTPC-Au multiplicity events from our data are shown in Fig.~\ref{fig:deta}(c). A constant fit to the near- and away-side difference gives a $\chi^2/{\rm ndf}=50/9$ and $6.4/9$, respectively, while a Gaussian fit to the near side gives $\chi^2/{\rm ndf}=2.3/8$. 
These differences resemble jetlike correlation features, consistent with a Gaussian peak on the near side and a uniform distribution on the away side. They therefore suggest that the difference is likely of jetlike origin. 

As a first attempt to ``address'' the jetlike correlated yield difference, the jetlike ratio parameter $\alpha$ is applied as a scaling factor to the low-activity data before it is subtracted from the high-activity data. This procedure assumes that the away-side correlated yield scales with the near-side one, which is based on momentum conservation arguments.
The resulting subtracted data are shown in Fig.~\ref{fig:deta}(d).
The shape of the near-side difference is the result of 
subtracting a narrow Gaussian from a wide one of equal area offset by a pedestal.
On the away side, 
once the low-activity data are scaled up, the correlated yields are consistent between high- and low-activity collisions as shown by the open circles in Fig.~\ref{fig:deta}(d). 
%
This suggests that the away-side difference between high- and low-activity events may be primarily due to a difference in jetlike correlations.

As seen in Table~\ref{tab:yields}, the fit pedestal values of $C$ also shows dependence on event activity. 
Finite correlated yields above \zyam\ exist on the near side at large $\deta$, where the near-side jet contribution should be minimal. 
This large $\deta$ correlation data will be studied elsewhere \cite{STAR:2014ridge}.

\begin{figure}
\begin{center}
\includegraphics[width=0.4\textwidth]{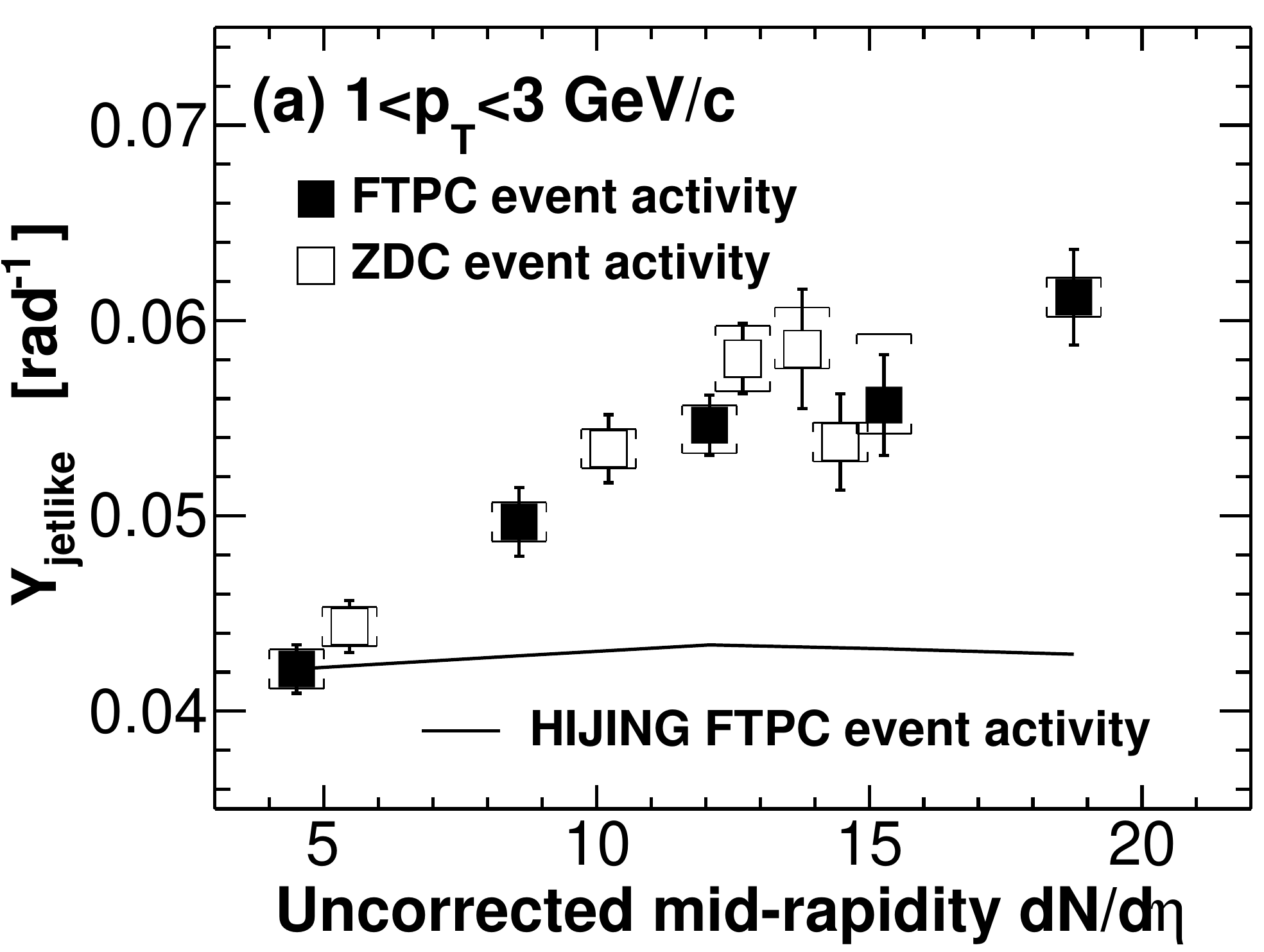}
\includegraphics[width=0.4\textwidth]{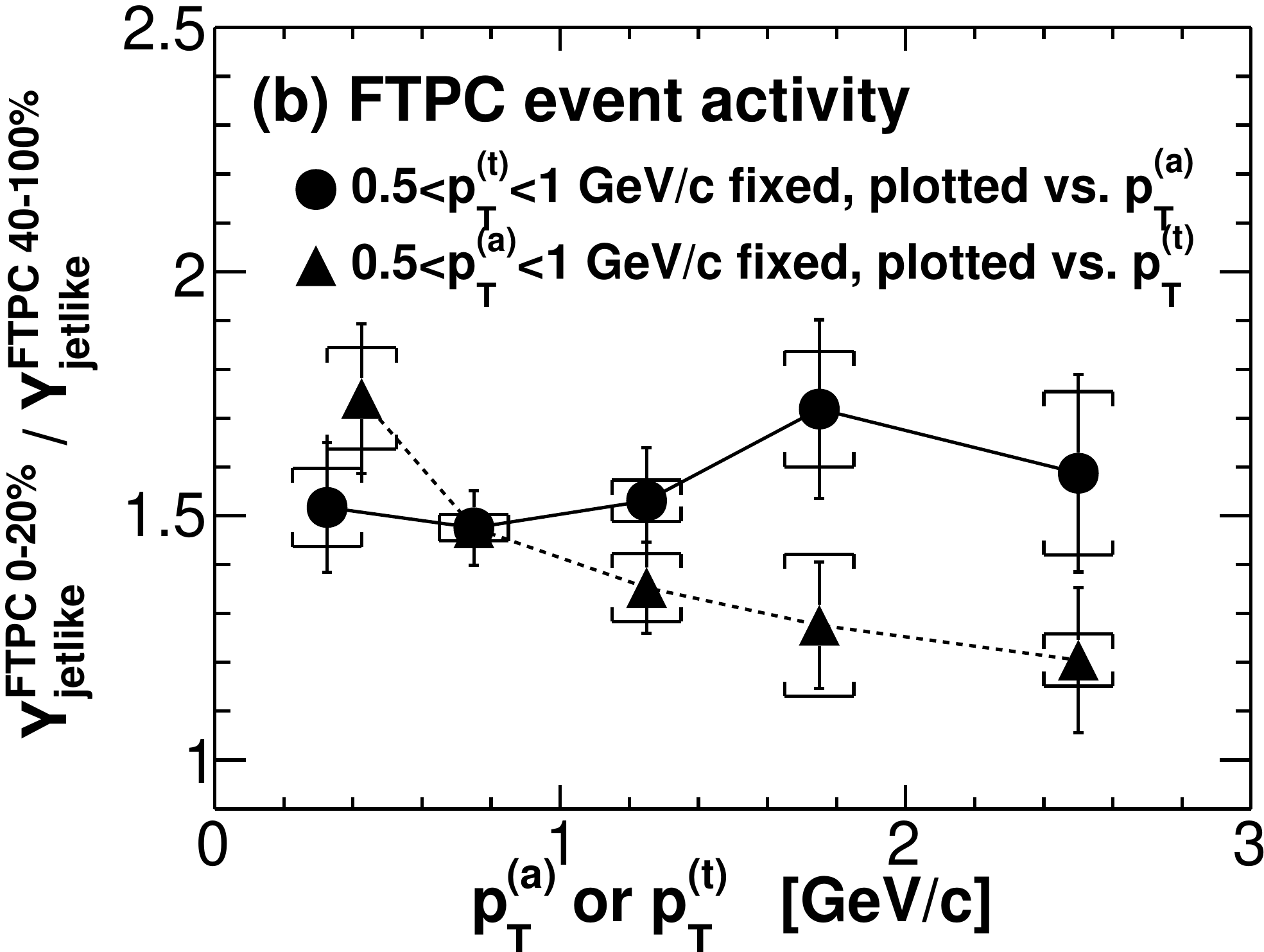}
\end{center}
\caption{(a) The near-side jetlike correlated yield obtained from Gaussian fit as in Fig.~\ref{fig:deta} as function of the uncorrected $dN/d\eta$ at midrapidity measured in the TPC. Two event selections are used: FTPC-Au multiplicity (filled squares) and ZDC-Au energy (open squares). The curve is the result from a \hijing\ calculation. (b) The ratio of the correlated yields in high over low FTPC-Au multiplicity events as a function of $\pta$ ($\ptt$) where $\ptt$ ($\pta$) is fixed. Error bars are statistical and caps show the systematic uncertainties.}
\label{fig:yield}
\end{figure}

To investigate further the influence of event selection on jetlike correlations, Fig.~\ref{fig:yield}(a) shows $\Yjet$ as a function of the event activity, represented by the uncorrected charged hadron multiplicity $dN/d\eta$ at midrapidity, in events selected according to the FTPC-Au multiplicity (solid squares) and ZDC-Au neutral energy (open squares), respectively. Five event samples are selected by each measure, corresponding to 60-100\%, 40-60\%, 20-40\%, 10-20\%, and 0-10\%
events. The systematic uncertainties are obtained from Gaussian fits to the $\deta$ correlations, as in Fig.~\ref{fig:deta}, varied by the \zyam\ systematic uncertainties.
Figure~\ref{fig:yield} (a) shows that the near-side jetlike correlated yield has a smooth linear dependence on event activity. 
Qualitatively similar behaviour is also observed at the LHC~\cite{Abelev:2014mva}.
Such a dependence is not observed in the \hijing\ \cite{Gyulassy:1994ew} simulation of $d$+Au collisions at RHIC as illustrated by the curve in Fig.~\ref{fig:yield}(a). The \hijing\ calculations are scaled down such that the lowest multiplicity bin matches the real data. 
The multiplicity dependence of the jetlike yield is clearly different for the \hijing\ simulations. 

The jetlike ratio $\alpha$ parameter 
can quantify the effect of the event selection on jetlike correlations. Figure~\ref{fig:yield}(b) shows the $\pt$ dependence of the $\alpha$ parameter. The systematic uncertainties are given by \zyam\ uncertainties as in Fig.~\ref{fig:yield}(a). Two sets of data points are shown: one (solid circles) has the trigger $\pt$ fixed to $0.5<\ptt<1$~\gev\ and shows the $\alpha$ parameter as a function of the associated particle $\pta$ with bin of $0.5$ GeV/$c$. This trigger $\pt$ range is similar to
$0.5<p_{T}^{(t)}<0.75$ GeV/$c$ used by PHENIX~\cite{Adare:2013piz}. The $\alpha$ parameter is larger than unity and relatively insensitive to $\pta$ for this particular $\ptt$ choice. The other set of points (solid triangles) shows $\alpha$ as function of $\ptt$ with a fixed $\pta$ of $0.5<\pta<1$~\gev. In this case the $\alpha$ parameter decreases with $\ptt$.

There could be multiple reasons for the event-selection effects on jetlike correlations. One could be a simple selection bias due to auto-correlation: if the away-side jet contributes to the total FTPC-Au multiplicity, high FTPC-Au multiplicity events would preferentially select jets either of larger energy or happening to fragment into more particles. However, such an auto-correlation bias is not observed in the \hijing\ model implementation as clearly shown in Fig.~\ref{fig:yield}(a). Event-activity dependent sampling of jet energies could
also be caused by other physics origins; for example, there could be positive correlations between particle production from jets and from underlying events. The dependence of jetlike correlations at midrapidity on forward event activity could be driven by such mechanisms as initial-state $k_T$ effects or final-state jet modifications by possible medium formation~\cite{Adams:2005dq,Adcox:2004mh} in the small \dAu\ collision system. 

\begin{figure}
\begin{center}
\includegraphics[width=0.4\textwidth]{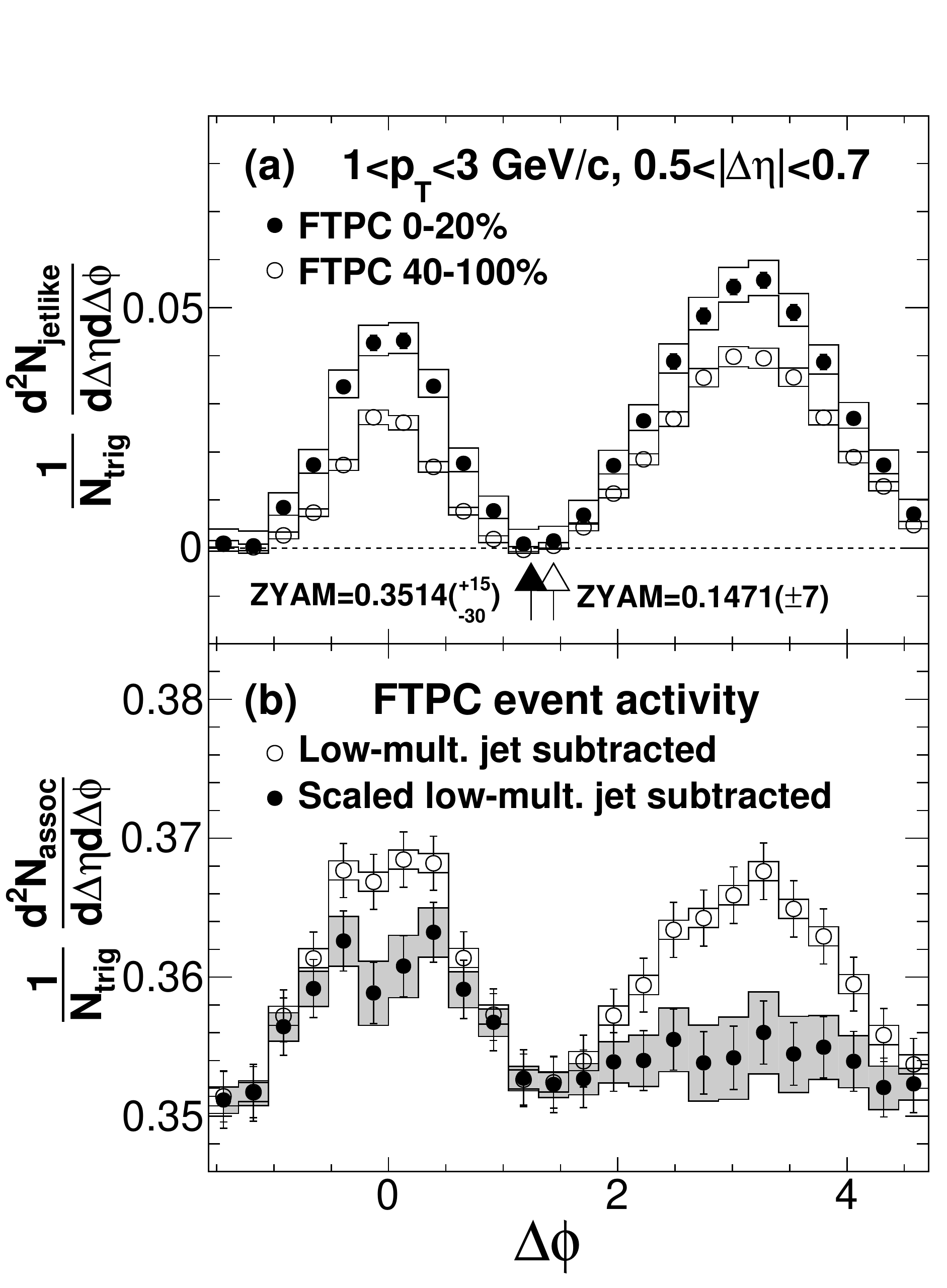}
\end{center}
\caption{(a) The dihadron correlated yield normalized per radian per unit of pseudorapidity as a function of $\dphi$ in \dAu\ collisions at low (40-100\%, open circles) and high (0-20\%, closed circles) FTPC-Au multiplicities. Trigger and associated particles are $1<\pt<3$~\gev\ within $0.5<|\deta|<0.7$. ZYAM positions are indicated
with arrows. 
(b) The raw associated yield at high FTPC-Au multiplicity minus the unscaled (open circles) and scaled (closed circles) \zyam-subtracted correlated yields at low FTPC-Au multiplicity versus $\dphi$. 
Error bars are statistical and boxes indicate the systematic uncertainties. 
}
\label{fig:dphi}
\end{figure}

The PHENIX experiment reported a double-ridge difference in the dihadron $\dphi$ correlations between high- and low-activity events in the acceptance range $0.48<|\deta|<0.7$ with event activity defined by total charge in the BBC at $-3.9<\eta<-3$~\cite{Adare:2013piz}. Figure~\ref{fig:dphi}(a) shows the STAR data analyzed in a similar acceptance of $0.5<|\deta|<0.7$ 
for high and low-activity events defined by the FTPC-Au which has similar $\eta$ coverage as PHENIX's BBC. 
The systematic uncertainties shown by the histograms are the quadratic sum of those due to efficiency and \zyam, as well as the \zyam\ statistical error, because it is common for all $\dphi$ bins.
The correlated yields are larger in high- than in low-activity collisions on both the near and away side as previously discussed. 
The difference of the raw associated yield (i.e. no ZYAM subtraction) in high-activity events minus the jetlike correlated yield (i.e. with ZYAM subtraction) in low-activity events is shown in Fig.~\ref{fig:dphi}(b) by the open points.
The systematic uncertainties are the quadratic sum of the statistical and systematic uncertainties on \zyam\ of the low-activity data. The additional 5\% efficiency uncertainty is not shown because it is an overall scale not affecting the shape of the dihadron correlation, therefore not affecting the physics conclusions.
Back-to-back double ridges are apparent and are qualitatively consistent with the PHENIX observation~\cite{Adare:2013piz}. However, the double-ridge structure is largely due to the residual jetlike correlation difference as demonstrated by our data above. Interpreting the double ridges as solely due to non-jet contributions in high-activity data is therefore premature.

Again, to account for the jetlike correlation difference, one may multiply the \zyam-subtracted low-activity data by the jetlike ratio $\alpha$ parameter before subtraction. 
Figure~\ref{fig:dphi}(b) shows, as the solid points, the raw associated particle yield (i.e. no ZYAM subtraction) in the high FTPC-Au multiplicity data after subtracting the $\alpha$-scaled 
jetlike
correlated yield (i.e. with ZYAM subtraction) in the low-multiplicity data.
The systematic uncertainties include the propagated total error from \zyam\ as well as the fit error on $\alpha$.
The near-side difference is non-zero above the underlying event baseline for the $\deta$ range used. This is because this simple $\alpha$ scaling does not account for the observed broadening of the near-side jetlike peak from low- to high-activity collisions, although the jetlike yield difference has been taken care of. This causes a significantly larger difference in the intermediate range of $0.5<|\deta|<0.7$. When $\deta$ range closer to zero is used, e.g.~$|\deta|<0.3$, the jetlike difference 
is dipped (below the baseline) on the near side after $\alpha$ scaling. 
This is shown by the negative solid data points at $\deta\sim0$ in Fig.~\ref{fig:deta}(d).
Barring from the difference caused by the broadening, there is a finite pedestal value from the near-side Gaussian+pedestal fit that increases with event activity as aforementioned. This pedestal difference remains in the near-side peak in Fig.~\ref{fig:dphi}(b).

After the jetlike contribution is removed by the scaled subtraction, the away-side difference is significantly diminished. The results are similar using the ZDC-Au event activity. This suggests that any possible contribution from non-jetlike long-range correlations, such as the back-to-back ridge, is small. 
Although it does a better job of removing jetlike contributions than a simple subtraction of low-activity from high-activity data, the scaled subtraction may not completely remove the jetlike contributions.
This is so for two reasons. One, the away-side jetlike yield in a given $\pt$ range may not strictly scale with the near-side one between high- and low-activity collisions, depending on the details of dijet production and fragmentation. Two, the jetlike correlation shapes, being different on the near side, can also be different on the away side, e.g.~due to increasing $k_T$ broadening (or acoplanarity) with event activity. 

In summary, dihadron correlations are measured at midrapidity using the STAR TPC as function of the forward rapidity event activity in \dAu\ collisions at $\snn=200$~GeV. The event activity is classified by the measured FTPC-Au forward charged particle multiplicity or the ZDC-Au zero-degree neutral energy. 
The correlated yields are extracted by subtracting the estimated background using \zyam. It is found that the correlated yield is larger in high- than in low-activity collisions and the $\deta$-dependence of the observed yield difference resembles jetlike features, suggesting a jetlike origin. 
There could be multiple reasons for the difference, ranging from simple auto-correlation biases to physical differences between high- and low-activity \dAu\ collisions.
The away-side correlation difference is significantly diminished after scaling the low-activity data by the ratio of the near-side jetlike correlated yields. 
Our data demonstrate that the dihadron correlation difference between high- and low-activity events at RHIC is primarily due to jets. In \dAu\ collisions at RHIC such event-selection effects on jetlike correlations must be addressed before investigating possible non-jet correlations such as anisotropic flow. 

We thank the RHIC Operations Group and RCF at BNL, the NERSC Center at LBNL and the Open Science Grid consortium for providing resources and support. This work was supported in part by the Offices of NP and HEP within the U.S.~DOE Office of Science, the U.S.~NSF, the Sloan Foundation, the DFG cluster of excellence `Origin and Structure of the Universe' of Germany, CNRS/IN2P3, STFC and EPSRC of the United Kingdom, FAPESP CNPq of Brazil, Ministry of Ed.~and Sci.~of the Russian Federation, NNSFC, CAS, MoST, and MoE of China, GA and MSMT of the Czech Republic, FOM and NWO of the Netherlands, DAE, DST, and CSIR of India, Polish Ministry of Sci.~and Higher Ed., Korea Research Foundation, Ministry of Sci., Ed.~and Sports of the Rep.~Of Croatia, Russian Ministry of Sci.~and Tech, and RosAtom of Russia.

\bibliography{ref}

\begin{thebibliography}{30}
\expandafter\ifx\csname natexlab\endcsname\relax\def\natexlab#1{#1}\fi
\expandafter\ifx\csname bibnamefont\endcsname\relax
  \def\bibnamefont#1{#1}\fi
\expandafter\ifx\csname bibfnamefont\endcsname\relax
  \def\bibfnamefont#1{#1}\fi
\expandafter\ifx\csname citenamefont\endcsname\relax
  \def\citenamefont#1{#1}\fi
\expandafter\ifx\csname url\endcsname\relax
  \def\url#1{\texttt{#1}}\fi
\expandafter\ifx\csname urlprefix\endcsname\relax\def\urlprefix{URL }\fi
\providecommand{\bibinfo}[2]{#2}
\providecommand{\eprint}[2][]{\url{#2}}

\bibitem[{\citenamefont{Arsene et~al.}(2005)}]{Arsene:2004fa}
\bibinfo{author}{\bibfnamefont{I.}~\bibnamefont{Arsene}} \bibnamefont{et~al.}
  (\bibinfo{collaboration}{BRAHMS Collaboration}),
  \bibinfo{journal}{Nucl.Phys.} \textbf{\bibinfo{volume}{A757}},
  \bibinfo{pages}{1} (\bibinfo{year}{2005}), \eprint{nucl-ex/0410020}.

\bibitem[{\citenamefont{Back et~al.}(2005)}]{Back:2004je}
\bibinfo{author}{\bibfnamefont{B.}~\bibnamefont{Back}} \bibnamefont{et~al.}
  (\bibinfo{collaboration}{PHOBOS Collaboration}),
  \bibinfo{journal}{Nucl.Phys.} \textbf{\bibinfo{volume}{A757}},
  \bibinfo{pages}{28} (\bibinfo{year}{2005}), \eprint{nucl-ex/0410022}.

\bibitem[{\citenamefont{Adams et~al.}(2005{\natexlab{a}})}]{Adams:2005dq}
\bibinfo{author}{\bibfnamefont{J.}~\bibnamefont{Adams}} \bibnamefont{et~al.}
  (\bibinfo{collaboration}{STAR Collaboration}), \bibinfo{journal}{Nucl.Phys.}
  \textbf{\bibinfo{volume}{A757}}, \bibinfo{pages}{102}
  (\bibinfo{year}{2005}{\natexlab{a}}), \eprint{nucl-ex/0501009}.

\bibitem[{\citenamefont{Adcox et~al.}(2005)}]{Adcox:2004mh}
\bibinfo{author}{\bibfnamefont{K.}~\bibnamefont{Adcox}} \bibnamefont{et~al.}
  (\bibinfo{collaboration}{PHENIX Collaboration}),
  \bibinfo{journal}{Nucl.Phys.} \textbf{\bibinfo{volume}{A757}},
  \bibinfo{pages}{184} (\bibinfo{year}{2005}), \eprint{nucl-ex/0410003}.

\bibitem[{\citenamefont{Chatrchyan et~al.}(2013)}]{CMS:2012qk}
\bibinfo{author}{\bibfnamefont{S.}~\bibnamefont{Chatrchyan}}
  \bibnamefont{et~al.} (\bibinfo{collaboration}{CMS Collaboration}),
  \bibinfo{journal}{Phys.Lett.} \textbf{\bibinfo{volume}{B718}},
  \bibinfo{pages}{795} (\bibinfo{year}{2013}), \eprint{1210.5482}.

\bibitem[{\citenamefont{Abelev et~al.}(2013)}]{Abelev:2012ola}
\bibinfo{author}{\bibfnamefont{B.}~\bibnamefont{Abelev}} \bibnamefont{et~al.}
  (\bibinfo{collaboration}{ALICE Collaboration}), \bibinfo{journal}{Phys.Lett.}
  \textbf{\bibinfo{volume}{B719}}, \bibinfo{pages}{29} (\bibinfo{year}{2013}),
  \eprint{1212.2001}.

\bibitem[{\citenamefont{Aad et~al.}(2013)}]{Aad:2012gla}
\bibinfo{author}{\bibfnamefont{G.}~\bibnamefont{Aad}} \bibnamefont{et~al.}
  (\bibinfo{collaboration}{ATLAS Collaboration}),
  \bibinfo{journal}{Phys.Rev.Lett.} \textbf{\bibinfo{volume}{110}},
  \bibinfo{pages}{182302} (\bibinfo{year}{2013}), \eprint{1212.5198}.

\bibitem[{\citenamefont{Adams et~al.}(2005{\natexlab{b}})}]{Adams:2005ph}
\bibinfo{author}{\bibfnamefont{J.}~\bibnamefont{Adams}} \bibnamefont{et~al.}
  (\bibinfo{collaboration}{STAR Collaboration}),
  \bibinfo{journal}{Phys.Rev.Lett.} \textbf{\bibinfo{volume}{95}},
  \bibinfo{pages}{152301} (\bibinfo{year}{2005}{\natexlab{b}}),
  \eprint{nucl-ex/0501016}.

\bibitem[{\citenamefont{Abelev et~al.}(2009{\natexlab{a}})}]{Abelev:2009af}
\bibinfo{author}{\bibfnamefont{B.}~\bibnamefont{Abelev}} \bibnamefont{et~al.}
  (\bibinfo{collaboration}{STAR Collaboration}), \bibinfo{journal}{Phys.Rev.}
  \textbf{\bibinfo{volume}{C80}}, \bibinfo{pages}{064912}
  (\bibinfo{year}{2009}{\natexlab{a}}), \eprint{0909.0191}.

\bibitem[{\citenamefont{Alver et~al.}(2010)}]{Alver:2009id}
\bibinfo{author}{\bibfnamefont{B.}~\bibnamefont{Alver}} \bibnamefont{et~al.}
  (\bibinfo{collaboration}{PHOBOS Collaboration}),
  \bibinfo{journal}{Phys.Rev.Lett.} \textbf{\bibinfo{volume}{104}},
  \bibinfo{pages}{062301} (\bibinfo{year}{2010}), \eprint{0903.2811}.

\bibitem[{\citenamefont{Abelev et~al.}(2010)}]{Abelev:2009jv}
\bibinfo{author}{\bibfnamefont{B.}~\bibnamefont{Abelev}} \bibnamefont{et~al.}
  (\bibinfo{collaboration}{STAR Collaboration}),
  \bibinfo{journal}{Phys.Rev.Lett.} \textbf{\bibinfo{volume}{105}},
  \bibinfo{pages}{022301} (\bibinfo{year}{2010}), \eprint{0912.3977}.

\bibitem[{\citenamefont{Alver and Roland}(2010)}]{Alver:2010gr}
\bibinfo{author}{\bibfnamefont{B.}~\bibnamefont{Alver}} \bibnamefont{and}
  \bibinfo{author}{\bibfnamefont{G.}~\bibnamefont{Roland}},
  \bibinfo{journal}{Phys.Rev.} \textbf{\bibinfo{volume}{C81}},
  \bibinfo{pages}{054905} (\bibinfo{year}{2010}),
  \bibinfo{note}{erratum-ibid.~{\bf C82}, 039903 (2010)}, \eprint{1003.0194}.

\bibitem[{\citenamefont{Adare et~al.}(2013)}]{Adare:2013piz}
\bibinfo{author}{\bibfnamefont{A.}~\bibnamefont{Adare}} \bibnamefont{et~al.}
  (\bibinfo{collaboration}{PHENIX Collaboration}),
  \bibinfo{journal}{Phys.Rev.Lett.} \textbf{\bibinfo{volume}{111}},
  \bibinfo{pages}{212301} (\bibinfo{year}{2013}), \eprint{1303.1794}.

\bibitem[{\citenamefont{Bozek}(2011)}]{Bozek:2010pb}
\bibinfo{author}{\bibfnamefont{P.}~\bibnamefont{Bozek}},
  \bibinfo{journal}{Eur.Phys.J.} \textbf{\bibinfo{volume}{C71}},
  \bibinfo{pages}{1530} (\bibinfo{year}{2011}), \eprint{1010.0405}.

\bibitem[{\citenamefont{Bozek and Broniowski}(2013)}]{Bozek:2012gr}
\bibinfo{author}{\bibfnamefont{P.}~\bibnamefont{Bozek}} \bibnamefont{and}
  \bibinfo{author}{\bibfnamefont{W.}~\bibnamefont{Broniowski}},
  \bibinfo{journal}{Phys.Lett.} \textbf{\bibinfo{volume}{B718}},
  \bibinfo{pages}{1557} (\bibinfo{year}{2013}), \eprint{1211.0845}.

\bibitem[{\citenamefont{Dumitru et~al.}(2011)}]{Dumitru:2010iy}
\bibinfo{author}{\bibfnamefont{A.}~\bibnamefont{Dumitru}} \bibnamefont{et~al.},
  \bibinfo{journal}{Phys.Lett.} \textbf{\bibinfo{volume}{B697}},
  \bibinfo{pages}{21} (\bibinfo{year}{2011}), \eprint{1009.5295}.

\bibitem[{\citenamefont{Dusling and
  Venugopalan}(2013{\natexlab{a}})}]{Dusling:2012wy}
\bibinfo{author}{\bibfnamefont{K.}~\bibnamefont{Dusling}} \bibnamefont{and}
  \bibinfo{author}{\bibfnamefont{R.}~\bibnamefont{Venugopalan}},
  \bibinfo{journal}{Phys.Rev.} \textbf{\bibinfo{volume}{D87}},
  \bibinfo{pages}{054014} (\bibinfo{year}{2013}{\natexlab{a}}),
  \eprint{1211.3701}.

\bibitem[{\citenamefont{Dusling and
  Venugopalan}(2013{\natexlab{b}})}]{Dusling:2013oia}
\bibinfo{author}{\bibfnamefont{K.}~\bibnamefont{Dusling}} \bibnamefont{and}
  \bibinfo{author}{\bibfnamefont{R.}~\bibnamefont{Venugopalan}},
  \bibinfo{journal}{Phys.Rev.} \textbf{\bibinfo{volume}{D87}},
  \bibinfo{pages}{094034} (\bibinfo{year}{2013}{\natexlab{b}}),
  \eprint{1302.7018}.

\bibitem[{\citenamefont{Molnar et~al.}(2014)\citenamefont{Molnar, Wang, and
  Greene}}]{Molnar:2014mwa}
\bibinfo{author}{\bibfnamefont{D.}~\bibnamefont{Molnar}},
  \bibinfo{author}{\bibfnamefont{F.}~\bibnamefont{Wang}}, \bibnamefont{and}
  \bibinfo{author}{\bibfnamefont{C.~H.} \bibnamefont{Greene}}
  (\bibinfo{year}{2014}), \eprint{1404.4119}.

\bibitem[{\citenamefont{Adams et~al.}(2005{\natexlab{c}})}]{Adams:2004bi}
\bibinfo{author}{\bibfnamefont{J.}~\bibnamefont{Adams}} \bibnamefont{et~al.}
  (\bibinfo{collaboration}{STAR Collaboration}), \bibinfo{journal}{Phys.Rev.}
  \textbf{\bibinfo{volume}{C72}}, \bibinfo{pages}{014904}
  (\bibinfo{year}{2005}{\natexlab{c}}), \eprint{nucl-ex/0409033}.

\bibitem[{\citenamefont{Adams et~al.}(2003)}]{Adams:2003im}
\bibinfo{author}{\bibfnamefont{J.}~\bibnamefont{Adams}} \bibnamefont{et~al.}
  (\bibinfo{collaboration}{STAR Collaboration}),
  \bibinfo{journal}{Phys.Rev.Lett.} \textbf{\bibinfo{volume}{91}},
  \bibinfo{pages}{072304} (\bibinfo{year}{2003}), \eprint{nucl-ex/0306024}.

\bibitem[{\citenamefont{Abelev et~al.}(2009{\natexlab{b}})}]{Abelev:2008ab}
\bibinfo{author}{\bibfnamefont{B.}~\bibnamefont{Abelev}} \bibnamefont{et~al.}
  (\bibinfo{collaboration}{STAR Collaboration}), \bibinfo{journal}{Phys.Rev.}
  \textbf{\bibinfo{volume}{C79}}, \bibinfo{pages}{034909}
  (\bibinfo{year}{2009}{\natexlab{b}}), \eprint{0808.2041}.

\bibitem[{\citenamefont{Ackermann
  et~al.}(2003{\natexlab{a}})}]{Ackermann:2002ad}
\bibinfo{author}{\bibfnamefont{K.}~\bibnamefont{Ackermann}}
  \bibnamefont{et~al.} (\bibinfo{collaboration}{STAR Collaboration}),
  \bibinfo{journal}{Nucl.Instrum.Meth.} \textbf{\bibinfo{volume}{A499}},
  \bibinfo{pages}{624} (\bibinfo{year}{2003}{\natexlab{a}}).

\bibitem[{\citenamefont{Adler et~al.}(2003)}]{Adler:2003sp}
\bibinfo{author}{\bibfnamefont{C.}~\bibnamefont{Adler}} \bibnamefont{et~al.},
  \bibinfo{journal}{Nucl.Instrum.Meth.} \textbf{\bibinfo{volume}{A499}},
  \bibinfo{pages}{433} (\bibinfo{year}{2003}).

\bibitem[{\citenamefont{Anderson et~al.}(2003)}]{Anderson:2003ur}
\bibinfo{author}{\bibfnamefont{M.}~\bibnamefont{Anderson}}
  \bibnamefont{et~al.}, \bibinfo{journal}{Nucl.Instrum.Meth.}
  \textbf{\bibinfo{volume}{A499}}, \bibinfo{pages}{659} (\bibinfo{year}{2003}),
  \eprint{nucl-ex/0301015}.

\bibitem[{\citenamefont{Ackermann
  et~al.}(2003{\natexlab{b}})\citenamefont{Ackermann, Bieser, Brady, Cebra,
  Draper et~al.}}]{Ackermann:2002yx}
\bibinfo{author}{\bibfnamefont{K.}~\bibnamefont{Ackermann}},
  \bibinfo{author}{\bibfnamefont{F.}~\bibnamefont{Bieser}},
  \bibinfo{author}{\bibfnamefont{F.}~\bibnamefont{Brady}},
  \bibinfo{author}{\bibfnamefont{D.}~\bibnamefont{Cebra}},
  \bibinfo{author}{\bibfnamefont{J.}~\bibnamefont{Draper}},
  \bibnamefont{et~al.}, \bibinfo{journal}{Nucl.Instrum.Meth.}
  \textbf{\bibinfo{volume}{A499}}, \bibinfo{pages}{713}
  (\bibinfo{year}{2003}{\natexlab{b}}), \eprint{nucl-ex/0211014}.

\bibitem[{\citenamefont{Ajitanand et~al.}(2005)}]{Ajitanand:2005jj}
\bibinfo{author}{\bibfnamefont{N.}~\bibnamefont{Ajitanand}}
  \bibnamefont{et~al.}, \bibinfo{journal}{Phys.Rev.}
  \textbf{\bibinfo{volume}{C72}}, \bibinfo{pages}{011902}
  (\bibinfo{year}{2005}), \eprint{nucl-ex/0501025}.

\bibitem[{\citenamefont{{STAR}}()}]{STAR:2014ridge}
\bibinfo{author}{\bibnamefont{{STAR}}}, \bibinfo{note}{manuscript in
  preparation}.

\bibitem[{\citenamefont{Abelev et~al.}(2014)}]{Abelev:2014mva}
\bibinfo{author}{\bibfnamefont{B.~B.} \bibnamefont{Abelev}}
  \bibnamefont{et~al.} (\bibinfo{collaboration}{ALICE Collaboration})
  (\bibinfo{year}{2014}), \eprint{1406.5463}.

\bibitem[{\citenamefont{Gyulassy and Wang}(1994)}]{Gyulassy:1994ew}
\bibinfo{author}{\bibfnamefont{M.}~\bibnamefont{Gyulassy}} \bibnamefont{and}
  \bibinfo{author}{\bibfnamefont{X.-N.} \bibnamefont{Wang}},
  \bibinfo{journal}{Comput.Phys.Commun.} \textbf{\bibinfo{volume}{83}},
  \bibinfo{pages}{307} (\bibinfo{year}{1994}), \eprint{nucl-th/9502021}.

\end{thebibliography}
\end{document}